\documentclass[epj]{svjour}
\usepackage{psfig}
\usepackage{times}
\begin{document}
\title{Studying the $\omega N$
elastic and inelastic cross section with
nucleons\thanks{supported by DFG, RFFI and  Forschungszentrum
J\"{u}lich.}}
\author{Ye.S. Golubeva\inst{1}, W.Cassing\inst{2},
L.A. Kondratyuk\inst{3}, A. Sibirtsev\inst{2} and M.
B\"uscher\inst{4}}
\institute{Institute for Nuclear Research, 60th
October Anniversary Prospect 7A, \\ 117312 Moscow, Russia
\and Institut f\"ur Theoretische Physik, Universit\"at Giessen, \\
D-35392 Giessen, Germany
\and Institute of Theoretical and
Experimental Physics, B.\ Cheremushkinskaya 25, \\ 117259 Moscow,
Russia \and Forschungszentrum J\"ulich, Institut f\"ur Kernphysik,
D-52425 J\"ulich, Germany}
\date{Received: date / Revised version: date}

\abstract{We explore the possibility to measure the elastic and
inelastic $\omega N$  cross section in
$p{+}d{\to}d{+}\omega{+}p_{sp}$ and $p{+}A$ reactions. Our studies
indicate that the elastic scattering cross sections can be
determined for $\omega$ momenta above 1 GeV/c in $p{+}d$ reactions
by gating on high proton spectator momenta whereas the $\omega N$
absorption cross section down to low relative $\omega$ momenta is
most effectively studied in $p{+}A$ reactions at beam energies
2.0--2.7 GeV.
}
\authorrunning{E. Golubeva et al.}
\titlerunning{Studying the $\omega{-}N$ elastic and
inelastic cross section with nucleons}

\maketitle

\section{Introduction}

The $\omega$-meson properties at finite nuclear density
have become of recent interest especially in the context of
dilepton studies~\cite{Prep,Ralf}, where one hopes to see a shift
of the meson pole e.g. at density $\rho_0{\approx}$0.16
fm$^{-3}$~\cite{Metag,Golub96,Golub97,Weidmann}. However, the
in-medium properties of the vector mesons - as reflected in their
spectral functions - are presently a matter of strong debate.
Whereas at high baryon density and/or temperature some scaling of
the meson poles with the scalar quark condensate ${<}\bar{q}q{>}$
might be adequate~\cite{Brown}, the properties of the mesons in a
hadron gas at low density are essentially determined by the
scattering amplitudes with the hadrons in the local
environment~\cite{Rapp}. By means of dispersion relations the real
and imaginary part of these scattering amplitudes -- in the low
density approximation -- directly convert to mean-field potentials
and scattering rates, respectively~\cite{Kondrat}. Thus it is of
profound interest to obtain more precise information especially
about the $\omega N$  cross sections to construct its spectral
function at low baryon density.

We recall that the strong coupling of the $\rho$-meson to nucleon
resonances leads to large inelastic cross sections with nucleons
at low relative momenta which in turn results in a large
collisional width at density $\rho_0$
\cite{Rapp,Kondrat,Weise1,Weise2,Weise3,Peters}. If this {\it
melting} of the $\rho$-meson will also occur for the $\omega$
meson is of major importance for dilepton experiments, that aim at
detecting moderate but clear peaks in the dilepton invariant mass
spectra. Though there are a couple of theoretical estimates for
the $\omega N$  cross
sections~\cite{Weise1,Weise2,Weise3,Ko,Friman,Lykasov} the
theoretical predictions widely differ such that related
experiments will have to clarify the situation.

Moreover, the amplitude for $\omega$-meson production in proton-
nucleon reactions close to threshold so far is scarcely known --
especially with a deuteron in the final state -- and of interest
in itself.

In this work we aim at contributing to the latter question by
investigating theoretically within the Multiple Scattering Monte
Carlo (MSMC) approach and transport model calculations, to what
extent we might obtain experimental
information from various reactions at existing accelerators that
can be performed in the near future. In Section 2 we will study in
detail the perspectives of the reaction
$p+d{\to}d{+}\omega{+}p_{sp}$ close to threshold. In Section 3
proton induced reactions on nuclei will be investigated within a
coupled channel transport approach to explore the production and
propagation of the $\omega$-meson in finite nuclei whereas Section
4 is devoted to a summary of our work.

\section{Effects from $\omega$  rescattering in the reaction
$p{+}d{\to}d{+}\omega{+}p_{sp}$}

\begin{figure}[htb]
\phantom{aa}\vspace{-22mm}\hspace{-9mm}
\psfig{file=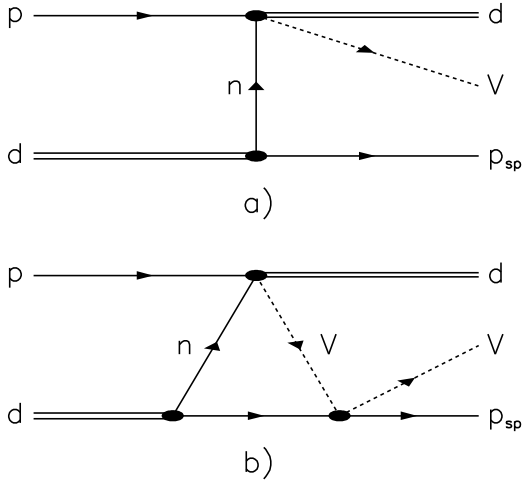,width=9.6cm} \vspace{-10mm} \caption{The
diagrams for vector meson ($V$) production in the $p{+}d{\to}d{+}V
+p_{sp}$ reaction without (a) and with $VN$ rescattering (b).}
\label{omeko6}
\end{figure}

In Ref.~\cite{Grishin99} we have considered the cross
section of the reaction $p{+}d{\to}d{+}\omega{+}p$
within the framework of the spectator
mechanism close to threshold energies (see diagram a)
in Fig.~\ref{omeko6}).
This mechanism gives the dominant contribution to the cross
section if the momentum of the proton-spectator from the deuteron
is below 100--150 MeV/c. The amplitude corresponding to the
diagram a) of Fig.~\ref{omeko6} can be written in
the following form in the
deuteron rest frame
\begin{equation}
\label{eq:Ma} M_a = f(pn \to d V) \int d^3 {\bf r} \ \exp(-i {\bf
p_2} {\bf r}) \ \phi_d({\bf r}),
\end{equation}
where $V$ is the vector meson ($\omega$), $\bf{p_2}$ is
the proton spectator momentum, $\phi_d({\bf r})$ is the deuteron
wave function and $f(pn{\to}dV)$ is the amplitude of the reaction
$pn{\to}dV$, related to the differential lab. cross section as
\begin{equation}
\label{eq:ds/do} \frac{d\sigma(pn \to dV) }{d\Omega}  = |f(pn \to
d V)|^2 .
\end{equation}

The produced vector meson can rescatter elastically on the
proton-spectator, thus transferring \, momenta larger than 200 MeV/c.
Near threshold the momentum of the produced $\omega$ meson  is quite
large: about 0.8-1 GeV/c. Therefore its wave length is much smaller than
the average distance between two nucleons in a deuteron. In this
case we can describe the rescattering amplitude (see diagram b) in
Fig.~\ref{omeko6}) within the framework of the eikonal approximation (cf.
Ref.~\cite{Kondrat3})
\begin{equation}
 \label{eq:Mb}
M_b = - f(pn \to d V) \int d^3 {\bf r} \ \exp(-i {\bf p_2} {\bf
r})\  \Theta(z) \Gamma({\bf b}) \phi_d({\bf r}),
\end{equation}
where the $z$- axis is directed along the V-meson momentum. The
elastic $Vp$ - scattering amplitude is related to the profile
function $\Gamma ({\bf b })$ by the standard expression
\begin{equation}
\label{eq:fMp} f(Vp \to Vp) = \frac{1}{2\pi ik} \int d^2 {\bf b} \
\exp(-i {\bf q} {\bf b})\  \Gamma({\bf b})
\end{equation}
with $k$ denoting the lab. momentum of the V-meson and ${\bf q}$
the momentum transfer.

The probability to detect the deuteron at the solid angle $d\Omega
_d$ in coincidence with the spectator of momentum $\bf p _2$ is
related to the differential cross section as
\begin{equation}
\label{eq:ds/do5} \frac{d^5\sigma(pd \to d V p_2)}{d\Omega d^3
p_2}  =  |M_a + M_b|^2 .
\end{equation}
The probability for the produced meson to rescatter can be found
by integrating the term proportional to $|M_b|^2$,  which is equal
to
\begin{equation}
 \label{eq:W}
W = \sigma_{el}(V p\to V p) \int dz \ \Theta(z) |\phi_d({\bf b}
=0,z)|^2 .
\end{equation}

Now the momentum distribution of the spectators can be written in
the form
\begin{equation}
\label{eq:spec}
\frac{d N(p)}{p^2 d p}{=}N (|\psi_d({p})|^2{+}
 |\Psi_{resc}(p)|^2{+}Interference \, Term ),
\end{equation}
where the first term is the contribution of the spectator
mechanism, with $\psi_d(p)$ denoting the deuteron wave function in
momentum space, while  $N$ is the normalization factor. The second
term $|\Psi_{resc}(p)|^2$ appears due to the rescattering of all
particles in the final state on the proton-spectator added
incoherently with the probabilities defined by Eq.~(\ref{eq:W}).
The interference between the spectator and rescattering amplitudes
is only important in a narrow region of spectator momenta, where
both contributions are of the same order of magnitude. For a more
detailed discussion of this point see, e.g.
Refs.~\cite{Kondrat3,Shmatikov,Nikolaev,Kolybasov}).

In order to obtain closer information on the vector meson
rescattering we employ the Multiple Scattering Monte Carlo (MSMC)
approach. An earlier version of this approach, denoted as
Intra-Nuclear Cascade (INC) model, has been applied to the
analysis of $\eta$ and $\omega$ production in $\bar p A$ and $p A$
interactions in Refs.~\cite{Golub92,Golub93}, respectively.
Recently this version of the INC model has been extended to
incorporate in-medium modifications of the mesons
produced~\cite{Golub96,Golub97} in hadron-nuclear collisions. For
an explicit presentation of the production cross sections employed
we refer the reader to Ref.~\cite{Golub97} or the
review~\cite{Prep}.

\begin{figure}[h]
\vspace{-7mm}
\psfig{file=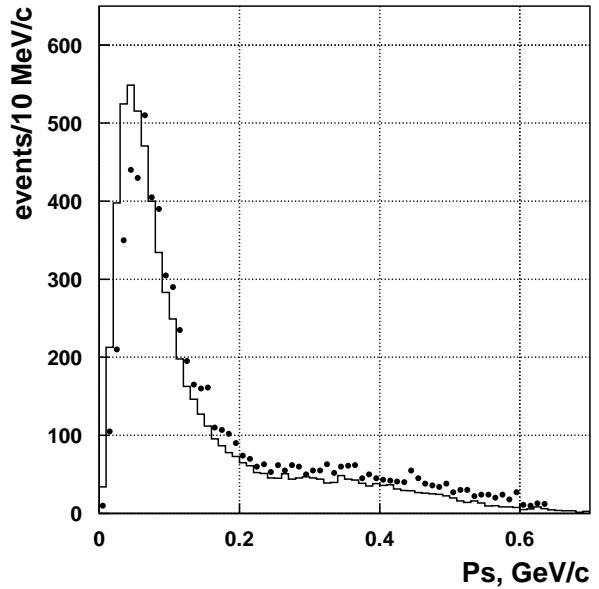,width=9.cm}
\caption{Momentum
distribution of proton-spectators in the reaction
$\bar{p}d{\to}3\pi^+2\pi^-p$. The
data are taken from Ref.~\cite{Ahmad} while the solid line is
calculated using Eq.~(\ref{eq:spec}).}
\label{fig:anpi}
\end{figure}

However, the INC model is valid only for medium and heavy nuclei
and cannot be directly applied to deuterons. In
order to perform simulations of scattering events in the case of
hadron -- deuteron interactions we use the Monte Carlo approach
for the description of single and double scattering terms. The
momentum distribution of a nucleon in the deuteron is described by
the Fourier transformed deuteron wave function squared
$|\psi_d({p})|^2$. The total probability of rescattering is
calculated for each secondary particle using Eq.(6). All  events
are generated using realistic angular distributions for the
elementary subprocesses. In the case of $\omega$  meson
rescattering this information is not available from data and we
employ different assumptions on its cross section, i.e. in
magnitude and angular distribution.

Experimentally, proton-spectator spectra have been measured in
$\bar{p}d$ annihilation~\cite{Bizzarri,Ahmad}
to some extent.  As an example -- and to testify the applicability
of the model to such type of reactions -- we present
in Fig.~\ref{fig:anpi} the calculated momentum distribution of
proton spectators in the reaction $\bar{p}d{\to}3\pi^+2\pi^-p$
at rest in comparison to the data from Ref.~\cite{Ahmad}.  These
data are easy to interpret: the sharp peak
with a maximum near 60 MeV/c corresponds to the spectator
mechanism and the long tail above 300 MeV/c is due to the
rescattering of pions. The solid line in Fig.~\ref{fig:anpi} is
calculated using Eq.~(\ref{eq:spec}) (without the interference
term), where the deuteron wave
function was taken in the parameterization of Ref.~\cite{Lacomb} with S-
and D-waves included. The dependence of the second term on the
momentum $p$ in the lab. system is determined by the kinematical
conditions (like the mass of the rescatterred particle) and by the
angular dependence of the pion-nucleon scattering amplitude. We
find that the spectrum calculated by the MSMC approach is in good
agreement with the experimental data.

\begin{figure}[h]
\vspace{-5.5cm} \psfig{file=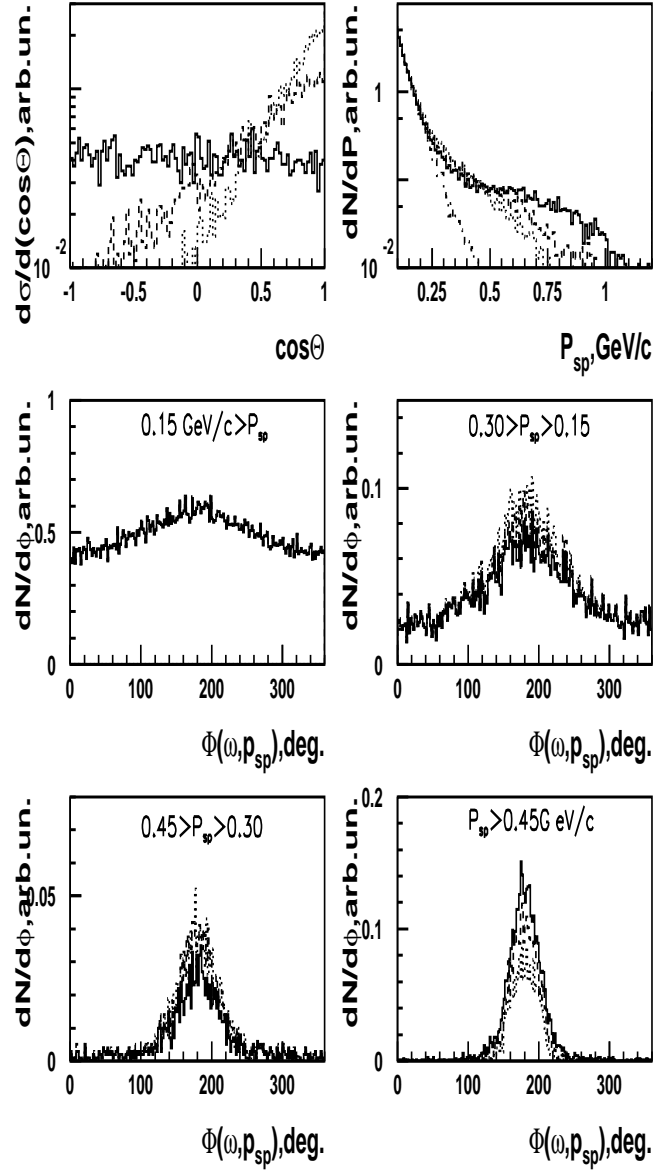,width=4.8cm,height=10.cm}
\vspace{11.cm} \caption{Angular distribution of $\omega N$
elastic rescattering (upper left) within 3 different models
indicated by the solid, dashed and dotted histograms (see text).
The resulting proton-spectator momentum distributions are shown in
the upper right part for the 3 models in the reaction
$p{+}d{\to}d{+}\omega{+}p_{sp}$. In all calculations the $\omega
N$ rescattering cross section $\sigma_{el}$=40 mb was assumed. The
distributions in the azimuthal angle $\Phi$ between the scattering
planes $p{-}p_{\omega}$ and $p{-}p_{sp}$ for spectators with
$p_{sp}{\leq}150$ MeV/c (middle left), $150{\leq}p_{sp}{\leq}300$
MeV/c (middle right), $300{\leq}p_{sp}{\leq}450$  (lower left) and
for fast spectators with $p_{sp}{\geq}450$ MeV/c (lower right) are
shown additionally.} \label{fig:romr1} \phantom{aa}\vspace{-5mm}
\end{figure}

The same approach was used to simulate the momentum distributions
of spectator protons in the reaction
$p{+}d{\to}d{+}\omega+p_{sp}$. Since the elastic rescattering
cross section of the $\omega$-meson is basically unknown and
especially its angular distribution at high relative momenta, we
study the observable effects within different models. In
Fig.~\ref{fig:romr1} we present three assumptions for the angular
distribution of $\omega N$ elastic rescattering (upper left).
These assumptions correspond to a parameterization of the
differential cross section $d\sigma /dt = C \exp(bt)$ with slope
parameters $b=0$, $6$ and 12~GeV$^{-2}$, respectively.

The resulting spectator momentum distributions are shown in the
upper right part of Fig.~\ref{fig:romr1}. The lowest (dotted)
histogram describes the contribution of the spectator term when
rescattering is absent. In this case the proton momentum spectrum
only extends to 450 MeV/c. The most pronounced effect from
rescatterings is seen for the isotropic model (solid histogram)
where the proton momentum spectrum extends to 1 GeV/c.

Furthermore, the distributions in the
azimuthal angle $\Phi$ between the scattering planes $p{-}p_\omega$
and $p{-}p_{sp}$ for spectators
with $p_{sp}{\leq}$150 MeV/c \, (middle left),
150${\leq}p_{sp}{\leq}$300 MeV/c \, (middle right),
300${\leq}p_{sp}{\leq}$ 450 MeV/c (lower left)
and for fast spectators with $p_{sp}{\geq}$
450 MeV/c (lower right), show the most pronounced rescattering effects
for fast spectators in line with the upper right part of the figure.

It is clearly seen that the spectator mechanism dominates at
$p_{sp}{\leq}300$ MeV/c while at $p_{sp}{\geq}
450$ MeV/c the main contribution stems from the rescattering term.
The best way to select the rescattering term is to look for the
correlation between the azimuthal angles of the two scattering
planes $p{-}p_{\omega}$ and $p{-}p_{sp}$ (lower figures).
This correlation is much more pronounced for fast spectators
($p_{sp}{\geq}450$ MeV/c) than for slower ones
($p_{sp}{\leq}300$ MeV/c).  It is worth to point out that
the slope of the spectator spectrum depends on the slope of the
$\omega{-}p$ angular distribution, while the correlation in the
angle $\Phi$ depends mainly on the value of the $\omega N$ elastic
cross section.

\begin{figure}[htb]
\vspace{-2.2cm}\hspace{2mm} \psfig{file=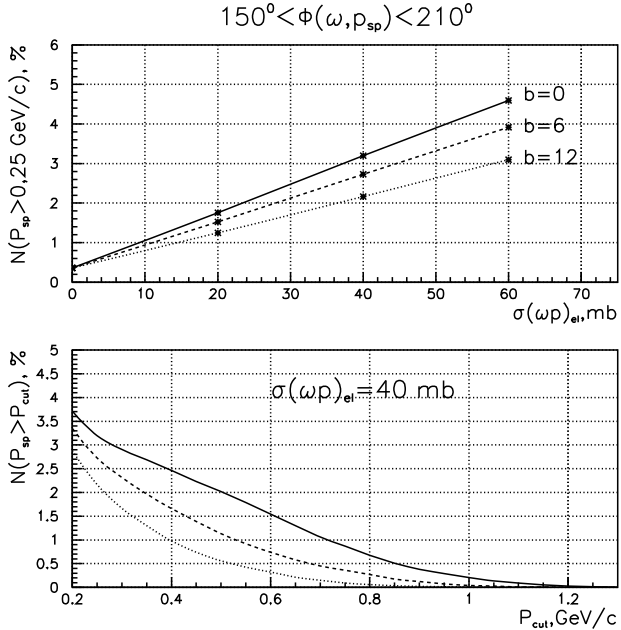,width=4.5cm}
\vspace{5.9cm} \caption{The number of events for the cut
$p_{sp}{\geq}250$ MeV/c and $150^0{<}{\phi}(\omega,
p_{sp}){<}210^o$ as a function of the elastic $\omega p$ cross
section for 3 different angular distributions (solid, dashed and
dotted lines) from the upper left part of
Fig.~\protect\ref{fig:romr1} is displayed in the upper part of the
figure. The lower part shows  the number of events as a function
of the cut in the proton-spectator momentum $p_{sp}$ for the 3
different angular distributions assuming an overall magnitude of
$\sigma_{el}(\omega p)$=40 mb in the reaction
$p{+}d{\to}d{+}\omega{+}p_{sp}$. The corresponding slope
parameters $b$ are given in GeV$^{-2}$(see text).}
\label{fig:romr3}
\end{figure}

Thus performing a cut for $p_{sp}{\geq}250$ MeV/c and $150^0 <
{\Phi(}\omega,p_{sp}){<}210^o$ one can determine essentially the
overall magnitude of the $\omega N$ elastic cross section (cf.
upper part of Fig.~\ref{fig:romr3}), whereas the number of events
as a function of the cut in the proton-spectator momentum $p_{sp}$
provides information on the $\omega{-}N$ angular distribution.
This is demonstrated in the lower part of Fig.~\ref{fig:romr3} for
the 3 models of the $\omega{-}p$ angular distributions, where the
solid, dashed and dotted lines correspond to the upper left part
of Fig.~\ref{fig:romr1}.

It is worth to point out that the relative contribution of the
rescattering term is about a few percent as compared to the
spectator term. In the case of small spectator momenta,  when the
spectator term is dominant, the reaction $p +
d{\to}d{+}\omega+p_{sp}$ can be identified by the missing mass
method measuring the forward deuteron in coincidence with a slow
spectator proton. We note that the spectator detection with thin
semiconductor counters has already successfully been applied  for
$\eta$-meson production in $pd$ collisions at
CELSIUS~\cite{Calen1,Calen2}. However, the selection of the
reaction $p + d{\to}d{+}\omega+p_{sp}$ with fast
($p_{sp}{\geq}250$ MeV/c) rescattered spectator protons is more
difficult since background particles (e.g. $\pi$-mesons) can be
misidentified as spectator protons. Thus, the direct detection of
the produced $\omega$-mesons by the detection of its decay
products like $\pi^0 \gamma$ is mandatory. We suggest to use a
neutral particle detector and to gate on three energetic photons,
where two photons should be correlated in their invariant mass to
the $\pi^0$ whereas the third photon should add to the invariant
mass of the $\omega$-meson according to its Dalitz decay
$\omega{\to}\pi^0{+}\gamma$ with a branching ratio of
$\simeq$8.5\%.

\begin{figure}[htb]
\vspace{-10mm}
\psfig{file=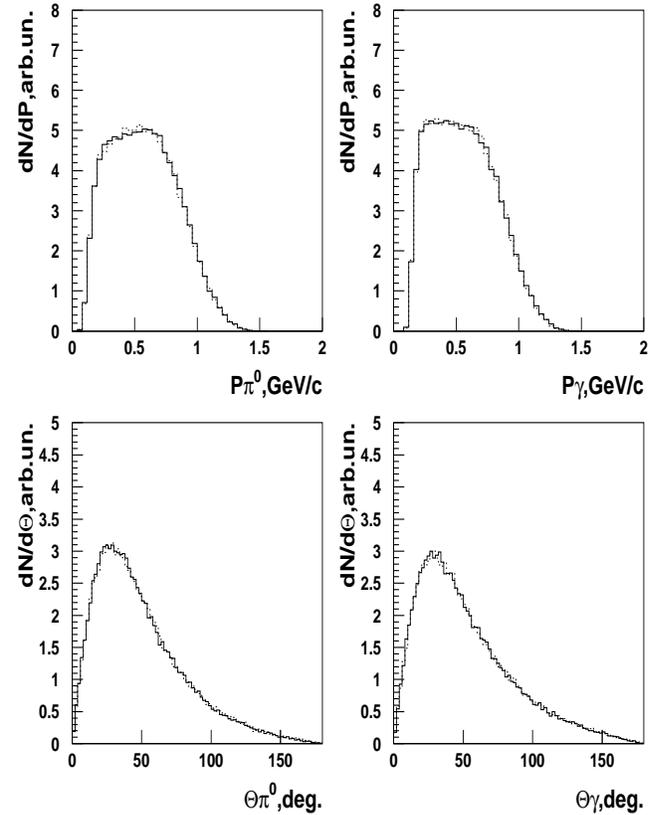,width=9.5cm,height=12.2cm}
\vspace{-1mm}
\caption{The momentum and angular distributions of $\pi^o$
mesons and photons from the reaction
$p{+}d{\to}d{+}\omega({\to}\pi^0\gamma){+}p_{sp}$ at
$T_p$=2 GeV. }
\label{fig:pf12}
\vspace{-5mm}
\end{figure}

The momentum and angular distributions of $\pi^0$-mesons
and photons in the reaction
$p{+}d{\to}d{+}\omega({\to}\pi^0\gamma){+}p_{sp}$ at $T_p$=2
GeV are shown in Fig.~\ref{fig:pf12}. The momentum spectra of
pions and photons have maxima at 0.5 GeV/c. Thus the detection of
such energetic photons will be a very good trigger for $\omega$
production, i.e. the selection of the reaction can be done by
measuring the $\omega$ with a fast forward deuteron in
coincidence. We note that such measurements  might well be
performed at COSY-J\"ulich.

\section{Proton induced reactions on nuclei}

The production of vector mesons in proton induced reactions on
light and heavy nuclei has been studied before for $\rho$-meson
propagation in the medium by looking at $\pi^+ \pi- $ invariant
mass spectra in Ref.~\cite{Sibirt98}. For the dynamical studies we
again employ the coupled channel transport model~\cite{Ehehalt}
with elementary production cross sections as described in
Refs.~\cite{Prep}. The transport approach~\cite{Ehehalt} allows to
account for the final state interactions (FSI) of all hadrons and
especially the elastic and inelastic $\omega N$ scatterings. For
the experimental detection of the $\omega$-meson we again suggest
to gate on three energetic photons.

In order to include the effects from $\omega N$ elastic
rescattering we employ the cross section from Ref.~\cite{Lykasov}
in the parameterization
\begin{equation}
\sigma_{el} = 5.4 + 10*\exp(-0.6 p_\omega) \ [\rm{mb}],
\end{equation}
where $p_\omega$ is the $\omega N$ relative laboratory momentum in
GeV/c. For the inelastic channels we have to consider various
reactions (cf. Ref.~\cite{Lykasov}). Among them the $\omega +
N{\to}\pi{+}N$ cross section can be determined quite well from the
available data on the exclusive $\omega$ production in $\pi N$
reactions from Ref.~\cite{LB} by exploiting detailed balance. The
$\omega N$ inelastic cross sections due to other reaction channels
have been calculated in Ref.~\cite{Lykasov} by a meson-exchange
model and, in principle, are determined within the uncertainties
of the coupling constants and the form factors employed. As a
guide we thus use the  inelastic $\omega N$ cross section
$\sigma_{inel}$ from Ref.~\cite{Lykasov} within the
parameterization
\begin{equation}
\label{sin}
\sigma_{inel} = 20 + 4.0/p_\omega \ [\rm{mb}],
\end{equation}
with $p_\omega$  given in GeV/c.

\begin{figure}[h]
\vspace{-7mm} \psfig{file=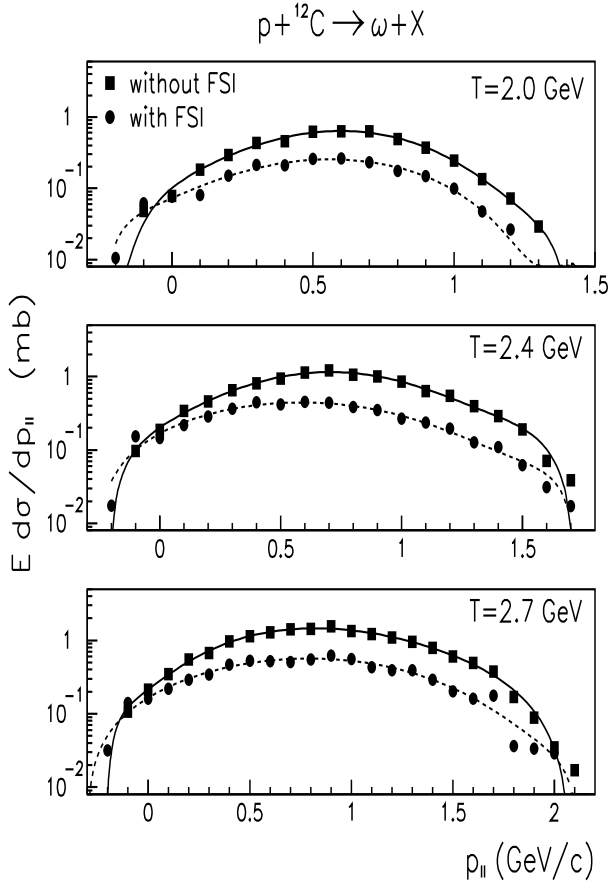,width=9.cm,height=13cm}
\vspace{-5mm} \caption{The longitudinal  momentum distribution of
$\omega$-mesons produced in  $p+^{12}C$ collisions at 2.0, 2.4 and
2.7 GeV. The symbols show the transport calculations while the
lines are drawn to guide the eye. The squares  show the
calculations without $\omega N$ final state interactions while the
full dots include FSI.} \label{omeko3}
\end{figure}

The momentum distribution of $\omega$-mesons in
beam direction,
\begin{equation}
E \frac{d \sigma}{dp_\parallel} = \int dp^2_\perp E \frac{d^3
\sigma}{d^3 p},
\end{equation}
is shown in Fig.~\ref{omeko3} for $p{+}^{12}C$ collisions at beam
energies of 2.0, 2.4 and 2.7 GeV. The dashed lines in
Fig.~\ref{omeko3} (full dots) show the calculations by taking into
account the elastic and inelastic $\omega N$ interactions, while
the solid lines indicate the results obtained without FSI (full
squares).

At all energies considered  the $\omega$ longitudinal momentum
distribution in the laboratory is extending over a broad momentum
regime from negative momenta even up  to $p_\parallel \approx$ 2.0
GeV/c at $T_{lab}$=2.7 GeV. By comparing the dashed and solid
lines we find a substantial reduction of the $\omega$-meson yield
at medium and high momenta even for $^{12}C$. Furthermore, for
$\omega$ momenta in the Fermi-momentum regime $p_F{\leq}$ 0.23
GeV/c a suppression is less obvious. This is due to elastic
${\omega}N{\to}{\omega}N$ scattering which essentially populates
low $p_\parallel$ in the laboratory system.

\begin{figure}[h]
\vspace{-5mm} \psfig{file=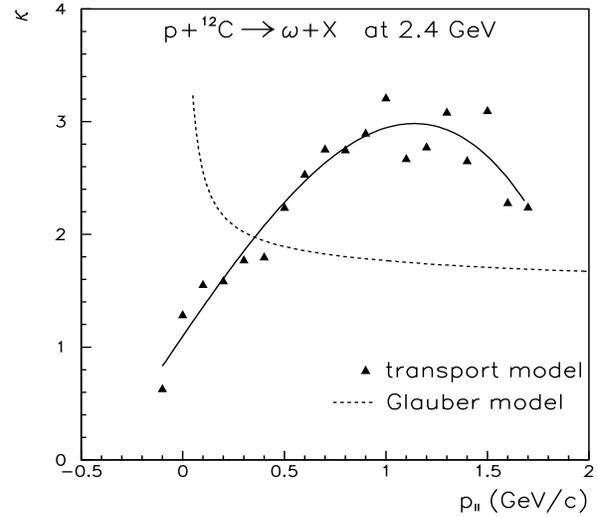,width=9.cm,height=8cm}
\vspace{-2mm} \caption{The ratio $\kappa$ of the $\omega$-meson spectrum
without FSI to the $\omega$-spectrum including FSI (triangles) from 
the transport calculation for $p + ^{12}C$ collisions at 2.4
GeV. The solid line is drawn to guide the eye while the dotted
line shows a Glauber calculation including the momentum dependent
$\omega N$  cross section.} \label{omeko5}
\end{figure}

The latter phenomenon is more clearly demonstrated in
Fig.~\ref{omeko5} where we display the ratio $\kappa$ of the
$\omega$ spectrum without FSI to the $\omega$ spectrum including
FSI (triangles) from the transport calculation for $p{+}^{12}C$ at
2.4 GeV. The solid line in Fig.~\ref{omeko5} is drawn to guide the
eye. Moreover, since the Glauber approximation is traditionally
considered~\cite{Glauber,Margolis1,Margolis2,Vercellin} as an
effective way to evaluate the cross section of  an unstable
particle with nucleons we perform a comparison between the Glauber
model  and the results from the transport approach.

We recall that within the Glauber model the A-dependence for
$\omega$-meson production and propagation in the nucleus is given
as~\cite{Sibirtsev4}
\begin{eqnarray}
A_{eff} =
\intop_{0}^{+\infty} \! bdb \intop_{-\infty}^{+\infty}\!\! \rho (b,z)dz
\intop_{0}^{2\pi}\! d\phi \nonumber \\
\times  [ \exp ( -{\sigma}_{pN} \!\!\intop_{-\infty}^{z}
\!\!\rho (b,\xi ) d \xi{-}{\sigma}_{\omega N}\!\!
\oint\rho ({\bf r}_\zeta) d\zeta ) ],
\label{aef}
\end{eqnarray}
where $\rho{(r)}$ is the single-particle density distribution
normalized to the target mass number. The last integration in
Eq.~(\ref{aef}) proceeds over the path of the produced
$\omega$-meson
\begin{equation}
r_{\zeta}^2 = (b+ \zeta cos\phi sin\theta)^2+(\zeta sin\phi sin\theta
)^2 +(z + \zeta cos\theta )^2 ,
\end{equation}
where $\theta$ is the emission angle of the $\omega$-meson
relative to the beam direction. Moreover, ${\sigma}_{pN}$ is the
$pN$ cross section while ${\sigma}_{\omega N}$ is the total
$\omega N$ cross section.

Within the Glauber model the $p_\parallel$-dependence of the
factor $\kappa$ is given by the ratio of  $A_{eff}$ -- calculated
with the momentum dependent $\omega N$ cross section~(\ref{sin})
-- to the result for ${\sigma}_{\omega N}$=0. The dashed line in
Fig.~\ref{omeko5} shows the factor $\kappa$ calculated by
Eq.~(\ref{aef}). It is clear that the calculations by the Glauber
model fail in describing the FSI as generated by the transport
calculation since the Glauber analysis discards elastic
rescattering and accordingly misses the population of the low
momentum part of the $\omega$ spectrum. Thus the factor $\kappa$
calculated by the transport model substantially differs from the
Glauber calculations at low $\omega$-meson momenta. Furthermore
the difference at $p_\parallel{>}$500 MeV/c is due to the fact that
the transport calculations account for the $\omega$ production by
secondary production processes, while the calculations by the
Glauber model do not.

\begin{figure}[h]
\phantom{aa}\vspace{-3mm}
\psfig{file=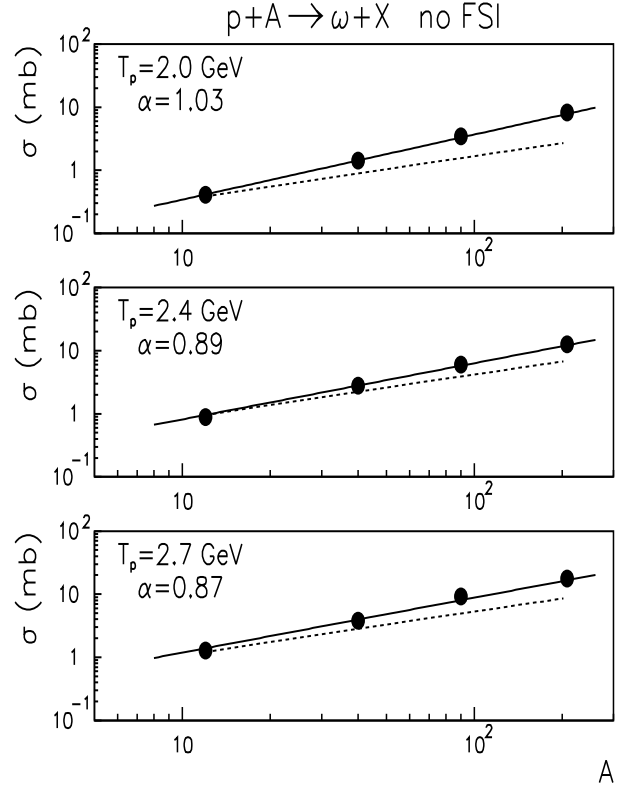,width=9.2cm,height=12.cm} \vspace{-8mm}
\caption{The $A$-dependence of $\omega$-meson production from
$p{+}A$ reactions at beam energy $T_p$=2.0, 2.4, and 2.7 GeV
calculated without  FSI. The full dots show the transport model
results while the solid lines are a fit by Eq.~(\ref{power}) with
parameters $\alpha$ as indicated in the figures. The dashed lines
show the calculations by the Glauber model without FSI, which
provide $\alpha{\approx}0.75$.} \label{omeko7}
\vspace{-5mm}
\end{figure}

Indeed, we note that at subthreshold energies in $p{+}A$ reactions
the secondary production channel with an intermediate pion
($p{+}N{\to}N{+}N{+}\pi$, $\pi{+}N{\to}\omega{+}N$) is the
dominant one according to the studies performed in
Refs.~\cite{Sibirtsev4,Honnef}. On the other hand, the
calculations by Eq.~(\ref{aef}) account only for the direct
$pN{\to}pN\omega$ production mechanism. Thus  the application of
the Glauber model is not appropriate in this case for an
experimental analysis.

The mass dependence of the total production cross section of
mesons can be exploited to provide some information on the
strength of the meson absorption cross section. In this respect we
show in Fig.~\ref{omeko7} (full dots) the expected mass
$A$-dependence of $\omega$ mesons from $p{+}A$ reactions at beam
energies $T_p$=2.0, 2.4, and 2.7 GeV as calculated by the
transport model without including FSI. The cross sections can be
well described at fixed energy by a power law
\begin{equation}
\label{power} \sigma (p+A\to \omega +X) \sim
A^{\alpha},
\end{equation}
where  the parameter $\alpha$ sheds some light on the reaction
mechanism. The solid lines in Fig.~\ref{omeko7} show a fit to the
transport model calculations with parameters $\alpha$ indicated in
the figure. Furthermore, for the direct $p{+}N{\to}\omega{+}p{+}N$
production (neglecting the $\omega$-meson FSI) we expect
$\alpha{\approx }0.75$ from the Glauber model (dashed lines in
Fig.~\ref{omeko7}) which, however, underestimates the mass
dependence obtained from the transport model calculations at all
energies and especially at $T_p$=2 GeV.

\begin{figure}[h]
\phantom{aa}\vspace{-5mm}
\psfig{file=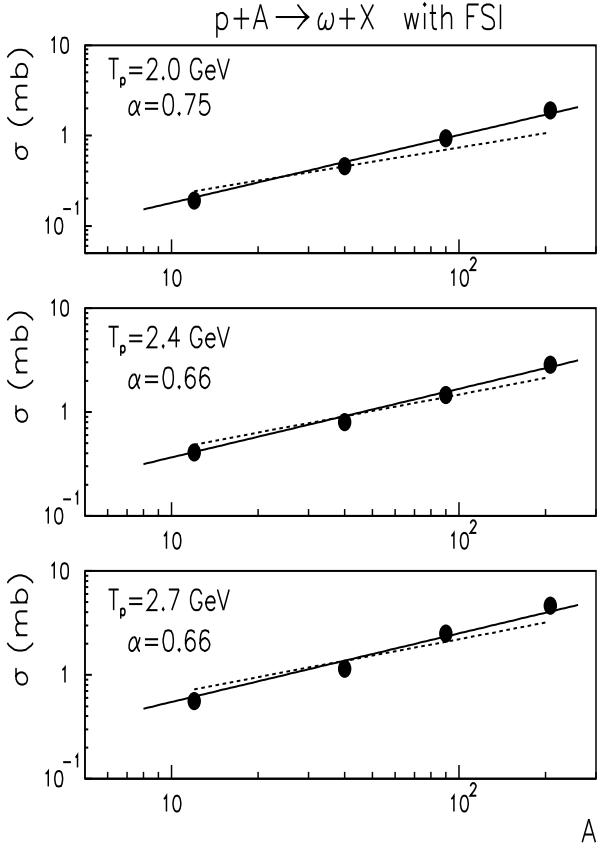,width=9.cm,height=12.2cm} \vspace{-2mm}
\caption{The $A$-dependence of $\omega$-meson production from
$p{+}A$ reactions at beam energy $T_p$=2.0, 2.4, and 2.7 GeV
calculated with $\omega N$ FSI. The full dots show the transport
model results while the solid lines are a fit by Eq.~(\ref{power})
with parameters $\alpha$ as indicated in the figures. The dashed
lines show the calculations by the Glauber model for
$\sigma_{\omega N}$=35 mb, which give $\alpha{\approx}0.55$.}
\label{omeko8}
\end{figure}

As discussed in Ref.~\cite{Sibirtsev4}-\cite{Sibirtsev} the higher
power in $A$ is due to secondary reaction channels involving an
intermediate pion, which are more frequent in heavy nuclear
targets and at lower energy. On the other hand, at bombarding
energies of 2.4 and 2.7 GeV the primary reaction channel dominates
such that we come closer to the results from the Glauber
calculation. Furthermore, the difference between the parameter
$\alpha$ obtained from the transport calculations to that from the
Glauber model at beam energies $T_p$=2.4 and 2.7 GeV actually
indicates the contribution from the secondary reaction mechanism.

The mass dependence of the $\omega$ production cross section
including the FSI as resulting from the transport calculations is
shown in Fig.~\ref{omeko8} in comparison to the Glauber model
employing a constant $\omega N$  cross section of 35 mb in the
latter (dashed lines). The respective power law parameters
$\alpha$ from the transport calculation differ with bombarding
energy and approach $\alpha$=2/3 above $\simeq$2.4 GeV bombarding
energy characterizing the surface dominance of $\omega$ meson
production in $p{+}A$ reactions due to strong absorption. On the
other hand, the Glauber model gives a rough description of the
mass dependence at higher energies for $\sigma_{\omega N}$=35 mb
(with $\alpha \approx$ 0.55).  This sensitivity thus might also be
explored experimentally, however, keeping in mind that the
measurements should be performed at different bombarding energies
to avoid systematic errors associated with different reaction
mechanisms.

\section{Summary}

In this work we have explored the possibility to measure the
elastic $\omega N$  cross section in
$p{+}d{\to}d{+}\omega{+}p_{sp}$ reactions and the inelastic
$\omega N$ cross section in $p{+}A$ collisions. Our studies based
on MSMC and transport calculations indicate that the elastic
scattering cross sections can be determined for $\omega$ momenta
above 1 GeV/c in $p+d$ reactions by gating on high proton
spectator momenta whereas the $\omega N$ absorption cross section
is most effectively studied in $p{+}A$ reactions around
$T_{lab}$=2.4 GeV via the mass dependence of the total production
cross section. Though the apparent power law in the target mass
$A$ is sensitive to the actual strength of the $\omega N$
inelastic cross section, its actual momentum dependence cannot be
extracted easily from experimental data due to strong effects from
elastic $\omega N$ scattering processes.

\section*{Acknowledgments}
We are grateful to  G. Lykasov and M. V. Rzjanin for helpful
discussions and access to their calculations prior to publication.
This work was  supported in part by INTAS grant No. 96-0597,
DFG, RFFI and  Forschungszentrum J\"{u}lich.

\end{document}